\documentclass{elsarticle}

\usepackage{graphicx}

\usepackage{amsmath}
\usepackage{amssymb}
\usepackage{latexsym}

\newproof{pf}{Proof}
\newdefinition{defn}[thm]{Definition}
\usepackage{threeparttable}
\usepackage{algorithm}
\usepackage{algorithmic}
\usepackage{subfigure}

\begin{document}

\begin{frontmatter}


\title{A Comparative Study of Discretization Approaches for Granular Association Rule Mining}
\author{Xu He}
\ead{hexu$\_$grclab@163.com}
\author{Fan Min\corref{cor1}}
\ead{minfanphd@163.com}
\author{William Zhu}
\ead{williamfengzhu@gmail.com}


\address{Lab of Granular Computing,
Zhangzhou Normal University, Zhangzhou 363000, China}

\cortext[cor1]{Corresponding author. Tel.: +86 133 7690 8359}

\begin{abstract}
Granular association rule mining is a new relational data mining approach to reveal patterns hidden in multiple tables.
The current research of granular association rule mining considers only nominal data.
In this paper, we study the impact of discretization approaches on mining semantically richer and stronger rules from numeric data.
Specifically, the Equal Width approach and the Equal Frequency approach are adopted and compared.
The setting of interval numbers is a key issue in discretization approaches,
so we compare different settings through experiments on a well-known real life data set.
Experimental results show that: 1) discretization is an effective preprocessing technique in mining stronger rules;
2) the Equal Frequency approach helps generating more rules than the Equal Width approach;
3) with certain settings of interval numbers, we can obtain much more rules than others.
\end{abstract}

\begin{keyword}
Granular association rule, discretization, Equal Width, Equal Frequency, relational data mining.
\end{keyword}
\end{frontmatter}

  %
  %
  \section{Introduction}\label{section: introduction}

Relational data mining schemes \cite{DzeroskiS2003Multi, DzeroskiS2001Relational} look for patterns that include multiple tables in the database.
Some meaningful issues \cite{DehapseL1998Finding, Jensen2000Frequent, Goethals2008Mining, Kavurucu2009ILP, Goethals2010Mining} are undisputed more common and more challenging than their transcriptions on a single data table.
Recently, people focus on the tasks of association rule and computing with granules \cite{Lin98Granular,YaoYao03Information,Yao00Granular,Zadeh97Towards,ZhuWang03Reduction,yao2004partition}.

Granular association rule mining \cite{MinHuZhu12GranularTwo,MinHuZhu12GranularFour} is a new approach to reveal patterns hidden in multiple tables.
This approach generates rules with four measures to reveal connections between concepts in two universes.
We consider a database with two entities \texttt{customer} and \texttt{product} connected by a relation \texttt{buys}.
An example of granular association rules might be ``40\% men like at least 30\% kinds of alcohol; 45\% customers are men and 6\% products are alcohol."
Here 45\%, 6\%, 40\% and 30\% are the \emph{source coverage}, the \emph{target coverage}, the \emph{source confidence} and the \emph{target confidence}, respectively. Numeric data are very common in real world problems. Unfortunately, only nominal data are considered in the original definition of granular association rule \cite{MinHuZhu12GranularTwo,MinHuZhu12GranularFour}.

We employed two discretization approaches, called the Equal Width approach and the Equal Frequency approach \cite{chiu1991information2,dougherty1995supervised}, to preprocess the numeric data.
The Equal Width approach confirms the minimum and maximum of the numeric data, and divides the range into $k$ equal-width discrete intervals.
The Equal Frequency approach confirms the minimum and maximum of the numeric data, and divides the range into $k$ intervals which have the same number of sorted values in ascending order.
Compare those two approaches by generated rules and candidates, we can obtain the strength one applied to granular association rule mining.

Experiments are undertaken on the publicly available MovieLens data set. We introduce two parameters $k_1$ and $k_2$. $k_1$ is the number of intervals for the age of the user, $k_2$ is the number of intervals for the released year of the movie.
The discretization approaches are implemented with Java in our open source software COSER (Cost sensitive rough set) \cite{Coser}.

Our experiment results show that discretization is effective preprocessing technique in mining stronger rules. The Equal Frequency and the Equal Width approach are both simple methods to discretize data, while achieving good results.
Given four measures thresholds, the Equal Frequency generates more rules than the other one.
For any pair of integers $(k_1,k_2)$, we can obtain a set of rules.
Through comparing the number of all the sets of rules, we obtain certain settings of discrete interval numbers through different approaches. When setting $k_1$ range from 8 to 10 and $k_2$ range from 10 to 12 through the Equal Frequency approach, we can obtain much more rules than other settings.

The remainder of the paper is organized as follows.
Section \ref{section: granular-rules} reviews granular association rule.
Section \ref{section: gra-rule-numeric} presents granular association rules on numeric data, we might mine semantically richer and stronger rules.
In Section \ref{section: approaches}, we describe each discretization approach and discuss its suitability for granular association rule mining.
Experiments on the MovieLens data set \cite{movielens} are discussed in Section \ref{section: experiments}.
Finally, Section \ref{section: conclusion} presents the concluding remarks and further research directions.

  %
  %
  \section{Granular association rule}\label{section: granular-rules}
In this section, we revisit granular association rule \cite{MinHuZhu12GranularFour}. We analyse the definition, and four measures of such rule. Moreover, we introduce the basic design of granular association rule mining.

  %
  %
  \subsection{The data model}\label{subsection: data-model}
First of all, we introduce the data model which is built on information systems and binary relations.
\begin{defn}\label{definition: ins}
$S = (U, A)$ is an information system, where $U = \{x_1, x_2, \dots, x_n\}$ is the set of all objects, $A = \{a_1, a_2, \dots, a_m\}$ is the set of all attributes, and $a_j(x_i)$ is the value of $x_i$ on attribute $a_j$ for $i \in [1..n]$ and $j \in [1..m]$.
\end{defn}

In an information system, any $A' \subseteq A$ induces an equivalence relation \cite{Pawlak82Rough,SkowronStepaniuk94Approximation}
\begin{equation}\label{equation: equivalent-relation}
E_{A'} = \{(x, y) \in U \times U| \forall a \in A', a(x) = a(y)\},
\end{equation}
and partitions $U$ into a number of disjoint subsets called \emph{blocks}.
The block containing $x \in U$ is
\begin{equation}\label{equation: block-contain-x}
E_{A'}(x) = \{y \in U| \forall a \in A', a(y) = a(x)\}.
\end{equation}
From another viewpoint, a pair $C = (A', x)$ where $x \in U$ and $A' \subseteq A$ is called a \emph{concept}.
The \emph{extension} of the concept is
\begin{equation}\label{equation: extension-concept}
ET(C) = ET(A', x) = E_{A'}(x);
\end{equation}
while the \emph{intension} of the concept is the conjunction of respective attribute-value pairs, i.e.,
\begin{equation}\label{equation: concept}
IT(C) =  IT(A', x) = \bigwedge_{a \in A'}\langle a: a(x) \rangle.
\end{equation}
The \emph{support} of the concept is the size of its extension divided by the size of the universe, namely,
\begin{equation}\label{equation: support-quantitative-concept}
\begin{array}{llll}
support(C) &= support(A', x) &= support(\bigwedge_{a \in A'}\langle a: a(x) \rangle)\\
&= support(E_{A'}(x)) &= \frac{|ET(A', x)|}{|U|}\\
&= \frac{|E_{A'}(x)|}{|U|}.
\end{array}
\end{equation}

\begin{defn}\label{definition: binary-relation}
Let $U = \{x_1, x_2, \dots, x_n\}$ and $V = \{y_1, y_2, \dots, y_k\}$ be two sets of objects.
Any $R \subseteq U \times V$ is a binary relation from $U$ to $V$.
The neighborhood of $x \in U$ is
\begin{equation}\label{equation: relation}
R(x) = \{y \in V | (x, y) \in R\}.
\end{equation}
\end{defn}

If $U = V$ and $R$ is an equivalence relation, $R(x)$ is the equivalence class containing $x$.
From this definition we know immediately that for $y \in V$,
\begin{equation}\label{equation: relation-reverse}
R^{-1}(y) = \{x \in U | (x, y) \in R\}.
\end{equation}

A binary relation is more often stored in the database as a table with two foreign keys.
In this way the storage is saved.
For the convenience of illustration, here we represented it with an $n \times k$ boolean matrix.

With Definitions \ref{definition: ins} and \ref{definition: binary-relation}, we propose the following definition.
\begin{defn}\label{definition: m-m-er} \cite{MinHuZhu12GranularTwo}
A many-to-many entity-relationship system (MMER) is a 5-tuple $ES = (U, A, V, B, R)$, where $(U, A)$ and $(V, B)$ are two information systems, and $R \subseteq U \times V$ is a binary relation from $U$ to $V$.
\end{defn}


  %
  %
  \subsection{Granular association rule with four measures}\label{subsection: rules-measures}
Now we come to the central definition of granular association rules.
\begin{defn} \cite{MinHuZhu12GranularTwo}
A \emph{granular association rule} is an implication of the form
\begin{equation}\label{equation: granular-association}
(GR): \bigwedge_{a \in A'}\langle a: a(x) \rangle \Rightarrow \bigwedge_{b \in B'}\langle b: b(y) \rangle,
\end{equation}
where $A' \subseteq A$ and $B' \subseteq B$.
\end{defn}

According to Equation (\ref{equation: support-quantitative-concept}), the set of objects meeting the left-hand side of the granular association rule is
\begin{equation}\label{equation: left-granular-rule}
LH(GR) = E_{A'}(x);
\end{equation}
while the set of objects meeting the right-hand side of the granular association rule is
\begin{equation}\label{equation: right-granular-rule}
RH(GR) = E_{B'}(y).
\end{equation}

The \emph{source coverage} of a granular association rule is
\begin{equation}\label{equation: source-coverage}
scoverage(GR) = |LH(GR)| / |U|.
\end{equation}
The \emph{target coverage} of $GR$ is
\begin{equation}\label{equation: target-coverage}
tcoverage(GR) = |RH(GR)| / |V|.
\end{equation}

There is a tradeoff between the source confidence and the target confidence of a rule.
Consequently, no values can be obtained directly from the rule.
To compute any one of them, we should specify the threshold of the other.
Let $tc$ be the target confidence threshold.
The \emph{source confidence} of the rule is
\begin{equation}\label{equation: source-confidence}
sconfidence(GR, tc)
= \frac{|\{x \in LH(GR) | \frac{|R(x) \cap RH(GR)|}{|RH(GR)|} \geq tc\}|}{|LH(GR)|}.
\end{equation}

Let $mc$ be the source confidence threshold, and
\begin{equation}\label{equation: K-boundary}
\begin{array}{ll}
|\{x \in LH(GR) | |R(x) \cap RH(GR)| \geq K + 1\}|\\
< mc \times |LH(GR)|\\
\leq |\{x \in LH(GR) | |R(x) \cap RH(GR)| \geq K\}|.
\end{array}
\end{equation}
This equation means that $mc \times 100\%$ elements in $LH(GR)$ have connections with at least $K$ elements in $RH(GR)$, but less than $mc \times 100\%$ elements in $LH(GR)$ have connections with at least $K + 1$ elements in $RH(GR)$.
The \emph{target confidence} of the rule is
\begin{equation}\label{equation: target-confidence}
tconfidence(GR, mc) = K / |RH(GR)|.
\end{equation}
In fact, the computation of $K$ is non-trivial.
First, for any $x \in LH(GR)$, we need to compute $tc(x) = |R(x) \cap RH(GR)|$ and obtain an array of integers.
Second, we sort the array in a descending order.
Third, let $k = \lfloor mc \times |LH(GR)|\rfloor$, $K$ is the $k$-th element in the array.

%
  %
  \subsection{Granular association rule mining}\label{subsection: mining}
The basic design of granular association rule mining is as follows.
\begin{defn}\label{definition: rule-mining}
The granular association rule mining.

\textbf{Input:} An $ES = (U, A, V, B, R)$, a minimal source coverage threshold $ms$, a minimal target coverage threshold $mt$, a minimal source confidence threshold $mc$, and a minimal target confidence threshold $tc$.

\textbf{Output:} All granular association rules satisfying $scoverage(GR) \geq ms$, $tcoverage(GR) \geq mt$, $sconfidence(GR) \geq mc$, and $tconfidence(GR) \geq tc$.
\end{defn}


 %
  %
  \section{Granular association rule on numeric data}\label{section: gra-rule-numeric}
There are many different types of data to describe objects.
Recently,  all data are implicitly considered to be nominal. However, in the real world applications, a very large proportion of data sets involve numerical data. One scheme to solve this problem is to divide numeric data into a number of intervals and regard each interval as a category. This process is usually named discrerization \cite{fayyad1993multi,bay2000multivariate,MinCaiLiuBai07Dynamic,MinLiuFang08Rough, MinCaiLiuBai07Dynamic,MinLiuFang08Rough}.
At present, the most important thing we intend to do is that we can mine semantically richer and stronger rules which cannot mine in primary data through discretization.
For instance, we give an information system in Table \ref{table: customer}, where $U$ = \{c1, c2, c3, c4, c5, c6, c7, c8, c9, c10\}, and $A$ = \{Age, Gender, Married, Salary\}. Among them, Age and Salary values are numeric data.
Another example is given by Table \ref{table: product}, where $U$ = \{p1, p2, p3, p4, p5, p6, p7, p8\}, and $A$ = \{Country, Category, Color, Price\}. Among them, Price values are numeric data.

A binary relation is more often stored in the database as a table with two foreign keys.
In this way the storage is saved.
For the convenience of illustration, here we represented it with an $n \times k$ boolean matrix.
An example is given by Table \ref{table: buys}, where $U$ is the set of customers as indicated by Table \ref{table: customer}, and $V$ is the set of products as indicated by Table \ref{table: product}.

\setlength{\tabcolsep}{16pt}
\begin{table}[tb]\caption{Customer}
\label{table: customer}
\begin{center}
\begin{tabular}{ccccc}
\hline
CID     &  Age    & Gender   &  Married     &  Salary  \\
\hline
c1      &  20     & Male     &  No          &  2000  \\
c2      &  25     & Female   &  Yes         &  2800  \\
c3      &  23     & Male     &  No          &  3500  \\
c4      &  26     & Female   &  Yes         &  2400  \\
c5      &  32     & Male     &  Yes         &  5600  \\
c6      &  36     & Male     &  Yes         &  4200  \\
c7      &  39     & Male     &  Yes         &  5000  \\
c8      &  40     & Female   &  Yes         &  5000  \\
c9      &  35     & Female   &  Yes         &  3400  \\
~c10     &  34     & Male     &  Yes         &  3600  \\
\hline
\end{tabular}
\end{center}
\end{table}

\setlength{\tabcolsep}{14pt}
\begin{table}[tb]\caption{Product}
\label{table: product}
\begin{center}
\begin{tabular}{ccccc}
\hline
PID    &  Country    &  Category &  Color   &  Price\\
\hline
p1     &  China      &  Staple   &  Yellow  &  2.0  \\
p2     &  Australia  &  Staple   &  Black   &  4.0  \\
p3     &  China      &  Daily    &  White   &  5.5  \\
p4     &  China      &  Meat     &  Red     &  8.0  \\
p5     &  Australia  &  Meat     &  Red     &  18.0 \\
p6     &  China      &  Alcohol  &  Yellow  &  3.0  \\
p7     &  France     &  Alcohol  &  Yellow  &  5.0  \\
p8     &  France     &  Alcohol  &  White   &  16.5 \\
\hline
\end{tabular}
\end{center}
\end{table}

\setlength{\tabcolsep}{10pt}
\begin{table}[tb]\caption{Buys}
\label{table: buys}
\begin{center}
\begin{tabular}{ccccccccc}
\hline
CID$\diagdown$ PID &  p1     &  p2     &  p3    & p4      & p5      & p6    & p7   & p8 \\
\hline
c1      &  1      &  0      &  0     &  1      & 1       & 1      & 0       & 0 \\
c2      &  1      &  0      &  0     &  1      & 0       & 1      & 0       & 0 \\
c3      &  0      &  0      &  1     &  0      & 1       & 0      & 1       & 1 \\
c4      &  0      &  1      &  0     &  1      & 1       & 1      & 0       & 0 \\
c5      &  0      &  1      &  1     &  1      & 0       & 0      & 1       & 1 \\
c6      &  0      &  1      &  0     &  1      & 0       & 0      & 1       & 0 \\
c7      &  1      &  1      &  1     &  1      & 0       & 0      & 1       & 1 \\
c8      &  0      &  1      &  1     &  0      & 1       & 1      & 1       & 0 \\
c9      &  1      &  0      &  1     &  0      & 1       & 0      & 1       & 0 \\
~c10     &  1      &  0      &  1     &  0      & 1       & 0      & 1       & 1 \\
\hline
\end{tabular}
\end{center}
\end{table}

At present, we indicate all of the numeric data from the information systems. And then divide numeric data into a number of intervals and regard each interval as a category, as shown in Tables \ref{table: discretization customer}, \ref{table: discretization product}. From the MMER given by Tables \ref{table: buys}, \ref{table: discretization customer}and \ref{table: discretization product} we may obtain the following interesting rule.

\setlength{\tabcolsep}{16pt}
\begin{table}[tb]\caption{Discretization for Age and Salary}
\label{table: discretization customer}
\begin{center}
\begin{tabular}{ccccc}
\hline
CID     &  Age    & Gender   &  Married     &  Salary  \\
\hline
c1      &  [20,25)     & Male     &  No          &  [2000, 2900)  \\
c2      &  [25,30)     & Female   &  No          &  [2000, 2900)  \\
c3      &  [20,25)     & Male     &  No          &  [2900, 3800)  \\
c4      &  [25,30)     & Female   &  Yes         &  [2000, 2900)  \\
c5      &  [30,35)     & Male     &  Yes         &  [3800, 4700]  \\
c6      &  [35,40]     & Male     &  Yes         &  [2900, 3800)  \\
c7      &  [35,40]     & Male     &  Yes         &  [4700, 5600)  \\
c8      &  [35,40]     & Female   &  Yes         &  [4700, 5600)  \\
c9      &  [35,40]     & Female   &  Yes         &  [2900, 3800)  \\
~c10     &  [30,35)     & Male    &  Yes         &  [2900, 3800)  \\
\hline
\end{tabular}
\end{center}
\end{table}

\setlength{\tabcolsep}{14pt}
\begin{table}[tb]\caption{Discretization for Price}
\label{table: discretization product}
\begin{center}
\begin{tabular}{ccccc}
\hline
PID    &  Country    &  Category &  Color   &  Price\\
\hline
p1     &  China      &  Staple   &  Yellow  &  [2.0, 7.3)  \\
p2     &  Australia  &  Staple   &  Black   &  [2.0, 7.3)  \\
p3     &  China      &  Daily    &  White   &  [2.0, 7.3)   \\
p4     &  China      &  Meat     &  Red     &  ~[7.3, 12.7)   \\
p5     &  Australia  &  Meat     &  Red     &  ~~[12.7, 18.0] \\
p6     &  China      &  Alcohol  &  Yellow  &  [2.0, 7.3) \\
p7     &  France     &  Alcohol  &  Yellow  &  [2.0, 7.3) \\
p8     &  France     &  Alcohol  &  White   &  ~~[12.7, 18.0] \\
\hline
\end{tabular}
\end{center}
\end{table}

\noindent(Rule 1) $\langle\textrm{Gender: Male}\rangle$ $\Rightarrow \langle\textrm{Category: Alcohol}\rangle$.

\noindent(Rule 2) $\langle\textrm{Age: [30, 35)}\rangle \wedge \langle \textrm{Gender: Male}\rangle$
$\Rightarrow \langle\textrm{Category: Alcohol}\rangle$.

\noindent(Rule 3) $\langle\textrm{Married: Yes}\rangle$ $\Rightarrow \langle\textrm{Country: China}\rangle$.

\noindent(Rule 4) $\langle\textrm{Married: Yes}\rangle \wedge \langle \textrm{Salary: [4700, 5600]}\rangle$\\
~~~~~~\indent\indent$\Rightarrow \langle\textrm{Country: China}\rangle \wedge \langle\textrm{Price: [2.0, 7.3)}\rangle$.

Rule 1 can be read as ``men like alcohol."
Rule 2 can be read as ``men whose age is between 30 and 35 like alcohol."
Rule 3 can be read as ``Married people like products made in China."
Rule 4 can be read as ``Married people whose salaries are between 4700 and 5600, like products made in China, which prices are between 2.0 and 7.3."

From above we can come to a conclusion, we can mine semantically richer and stronger rules which cannot be mined in primary data through discretization, such as Rules 2, 4. Given the same four measures threshold, Rule 2 has a semantically richer rule than Rule 1, and Rule 4 has a richer rule than Rule 3.
A detailed explanation of Rule 4 might be ``60\% married people like at least 60\% products, which prices are between 2.0 and 7.3; 70\% customers are married people, 62.5\% products of all products which prices are between 2.0 and 7.3."

  %
  %
  \section{Discretization approaches}\label{section: approaches}
In this section, we introduce different discretization approaches,
which can divide the numeric data into different intervals and regard each interval as a category. Given four measures thresholds, we can mine different rules.
Since the number of intervals is a key issue in discretization approaches, we try to use some different settings of interval numbers to can obtain the suitable one. Then we can mine appropriate granule association rules.

In this paper, we adopt two discretization approaches, namely the Equal Width approach and the Equal Frequency approach.
The two approaches are both simple methods to discretize data and have often been used to produce nominal data from numeric ones.

  %
  %
  \subsection{The Equal Width approach}\label{subsection: Equal Width}
The Equal Width approach confirms the minimal value $a_{0}$ and the maximal value $a_{k}$ of the numeric data, and divides the range into $k$ equal-width discrete intervals. Here k is a parameter supplied by the user. The approach calculates the discretization width

\begin{equation}\label{equation: lambda}
\lambda= \frac{a_{k}-a_{0}}{k}.
\end{equation}

These values form the boundary set $\{a_{0}, a_{1}, ..., a_{i}, ..., a_{k-1}, a_{k}\} $ for $\{[a_{0}, a_{1}),$ $..., [a_{i-1}, a_{i}), ..., [a_{k-1}, a_{k}]\}$, $a_{i}=a_{0}+i\lambda$,
where $i=1, 2, ...,k$.
The approach is applied to each numeric data independently.
Finally, we obtain discretization data.

  %
  %
  \subsection{The Equal Frequency approach}\label{subsection: Equal Frequency}
The Equal Frequency approach confirms the minimal value $b_{0}$, the maximal value $b_{k}$ of the numeric data, and sorts the values from in ascending order. Here k is a parameter supplied by the user. Divide the range into $k$ of intervals in order that every interval involves the same number of sorted values, These values form the boundary set $\{b_{0}, b_{1}, b_{2}, ..., b_{k-1}, b_{k}\}$ for $\{[b_{0}, b_{1}), [b_{1}, b_{2}), ..., [b_{k-1}, b_{k}]\}$.

We set different interval number $k$ to divide the numeric data, and use different discretization approaches to produce different intervals. We know that more interval numbers, higher confidence of intervals, and lower coverage of intervals. Compare those intervals to get the suitable one for rule mining.
For example, Table \ref{table: product} shows that the value of Price range from 2.0 to 18.0.
Set $k = 3$, we get the price of p3 is between 2.0 and 7.3 with the Equal Width approach, while it is between 2.0 and 7.0 with the Equal Frequency approach.
Set $k = 4$, we get the price of p3 is between 2.0 and 6.0 with the Equal Width approach, while it is between 2.0 and 5.5 with the Equal Frequency approach.
Comparing those intervals, we obtain that take advantage of interval numbers and discretization approach is very important to produce suitable intervals for mining rule.


  %
  %
  \section{Experiments on a real world data set}\label{section: experiments}

    %
  %
  \subsection{A movie rating data set}\label{subsubsection: movie-data}
The MovieLens data set \cite{movielens} assembled by the GroupLens project is widely used in recommender systems (see, e.g., \cite{BalabanovicM1997Fab,Herlocker1999collaborative,ScheinA2002ColdStart,MinZhu12Parametric,MinZhu12ParametricCold}).
We downloaded the data set from the Internet Movie Database \cite{movielens}.
The data set contains 100,000 ratings (1-5) from 943 users on 1,682 movies, with each user rating at least 20 movies \cite{ScheinA2002ColdStart}.
In order to run our algorithm, we preprocessed the data set as follows.

\begin{enumerate}
\item{Remove movie names.
They are not useful in generating meaningful granular association rules.}
\item{Use release year instead of release date.
In this way the granule is more suitable.}
\item{Select the movie genre.
In the original data, the movie genre is multi-valued since one movie may fall in more than one genre.
For example, a movie can be both Animation and Children's.
Unfortunately, granular association rules do not support this type of data at this time.
Since the main objective of this work is to test compare the performances of algorithms, we use a simple approach to deal with this issue.
That is to sort movie genres according to the number of users they attract, and only keep the one highest priority for the current movie.
We adopt the following priority (from high to low): Comedy, Action, Thriller, Romance, Adventure, Children, Crime, Sci-Fi, Horror, War, Mystery, Musical, Documentary, Animation, Western, FilmNoir, Fantasy, Unkown.}
\end{enumerate}

Our database schema is as follows.
\begin{enumerate}
\item[$\bullet$]{User (\underline{userID}, age, gender, occupation)}
\item[$\bullet$]{Movie (\underline{movieID}, releaseYear, genre)}
\item[$\bullet$]{Rates (\underline{userID, movieID})}
\end{enumerate}

According to given intervals $[0, 18)$, $[18, 25)$, $[25, 30)$, $[30, 35)$, $[35, 45)$, $[45,56)$, $[56, \infty)$, the age of the user is discretized by the GroupLens project.
And then we use release decade instead of release date for the movies range from 1920s to 1990s.
As a result, a manual discretization setting is given to divide numeric data to obtain a finer granule. The setting would be used to compare with other discretization approaches.

  %
  %
  \subsection{Results}\label{subsection: results}
In this section, we try to answer the following problems through experimentation.

\begin{enumerate}
\item{Compared with the manual discretization setting to mine rules, Which approach outperform, the Equal Width approach or the Equal Frequency approach?}
\item{Whether we can mine much semantically richer rules through discretization?}
\item{What are the certain settings of discrete interval numbers for the numeric data?}
\end{enumerate}

We undertake three sets of experiments to answer the questions one by one.

 %
  %
  \subsubsection{The performance of discretization approaches}\label{subsubsection: performance-algorithms}

\begin{figure}[tb]
            \subfigure[]{
            \begin{minipage}[b]{2.7in}
            \centering
             \includegraphics[width=2.65in]{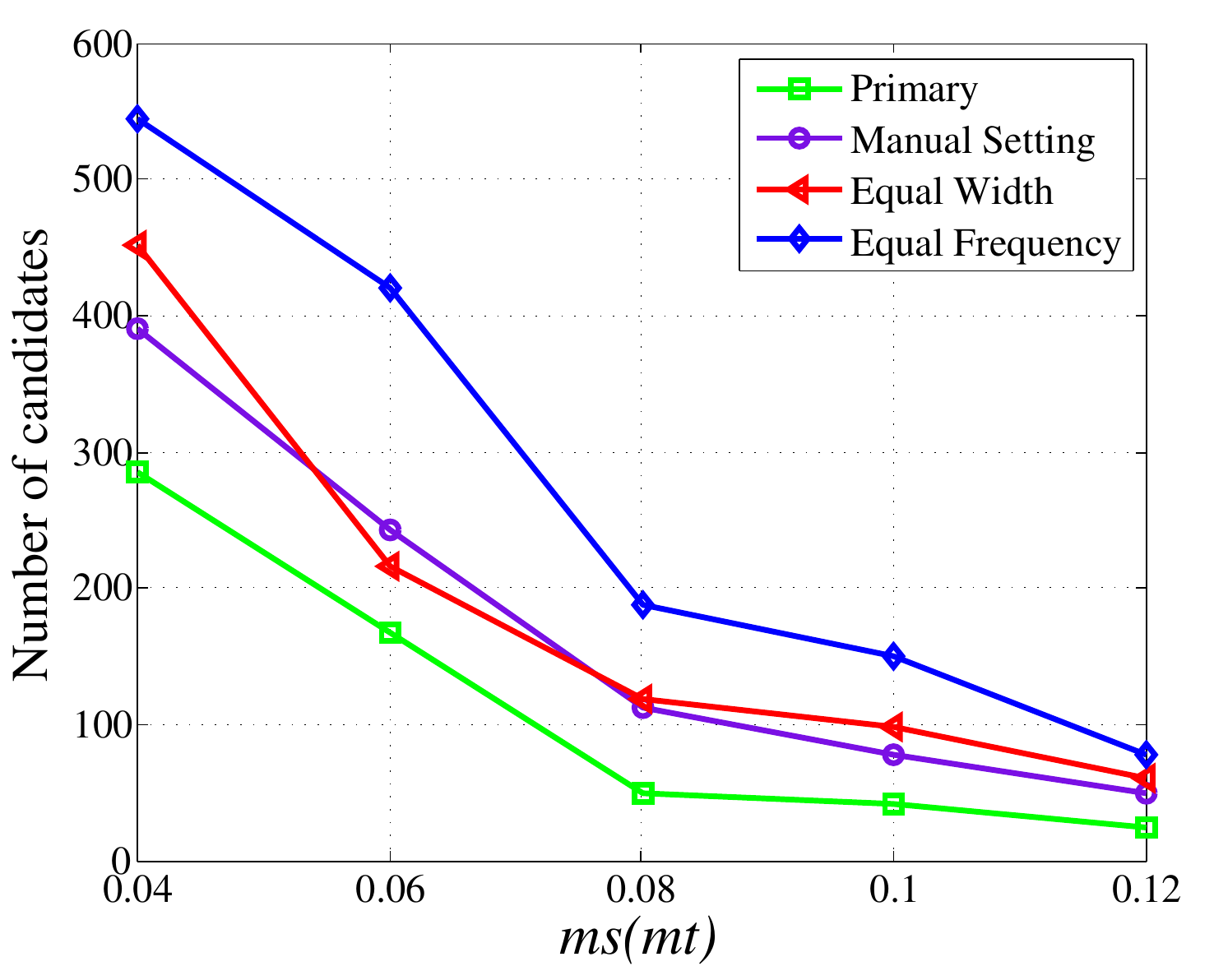}
            \end{minipage}
            }
            \subfigure[]{
            \begin{minipage}[b]{2.7in}
            \centering
             \includegraphics[width=2.65in]{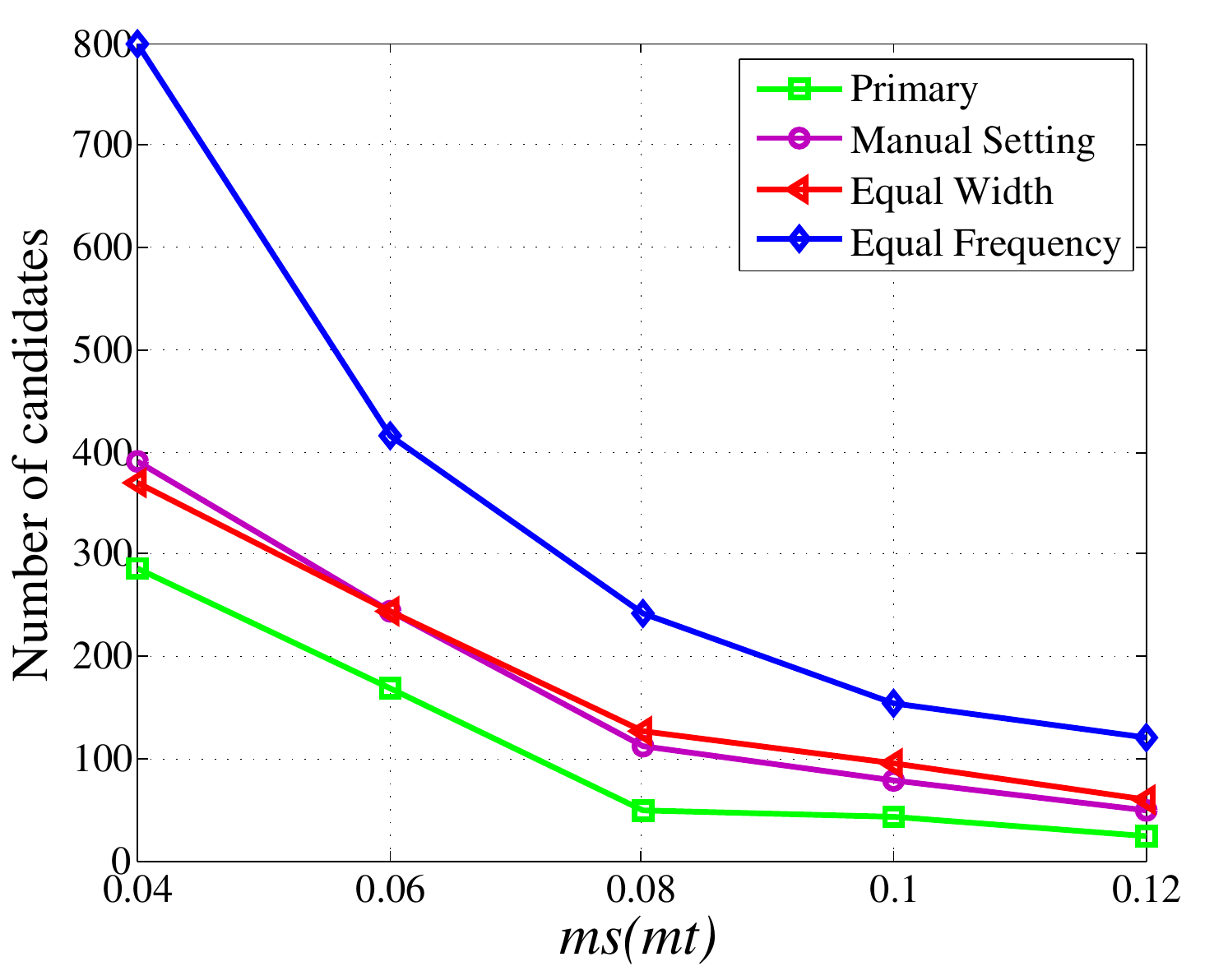}
            \end{minipage}
            }
            \caption{Number of candidates: (a) interval number $k = 4$; (b) interval number $k = 8$.}
            \label{figure: candidates-comp}
\end{figure}

\begin{figure}[tb]
            \subfigure[]{
            \begin{minipage}[b]{2.7in}
            \centering
             \includegraphics[width=2.65in]{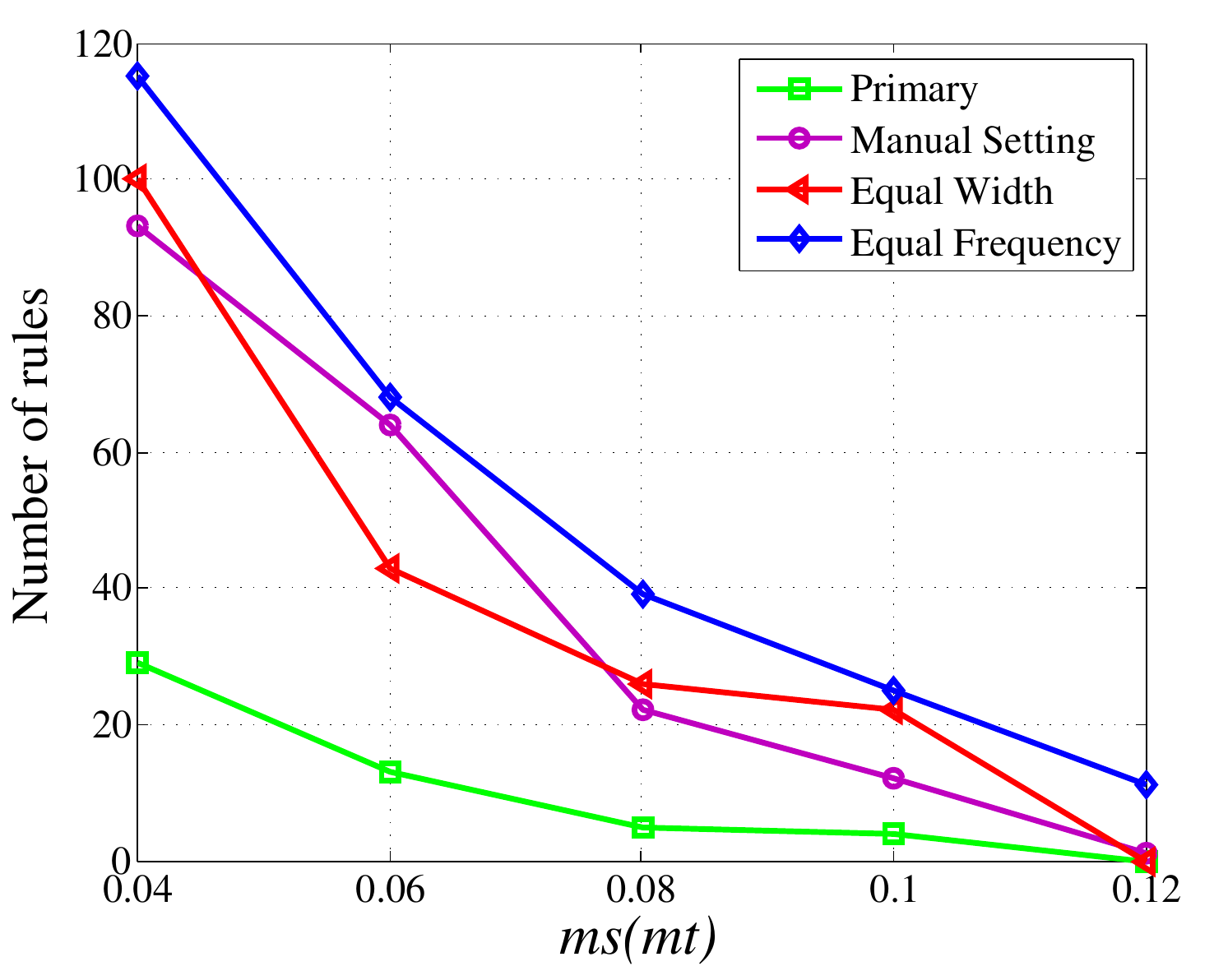}
            \end{minipage}
            }
            \subfigure[]{
            \begin{minipage}[b]{2.7in}
            \centering
             \includegraphics[width=2.65in]{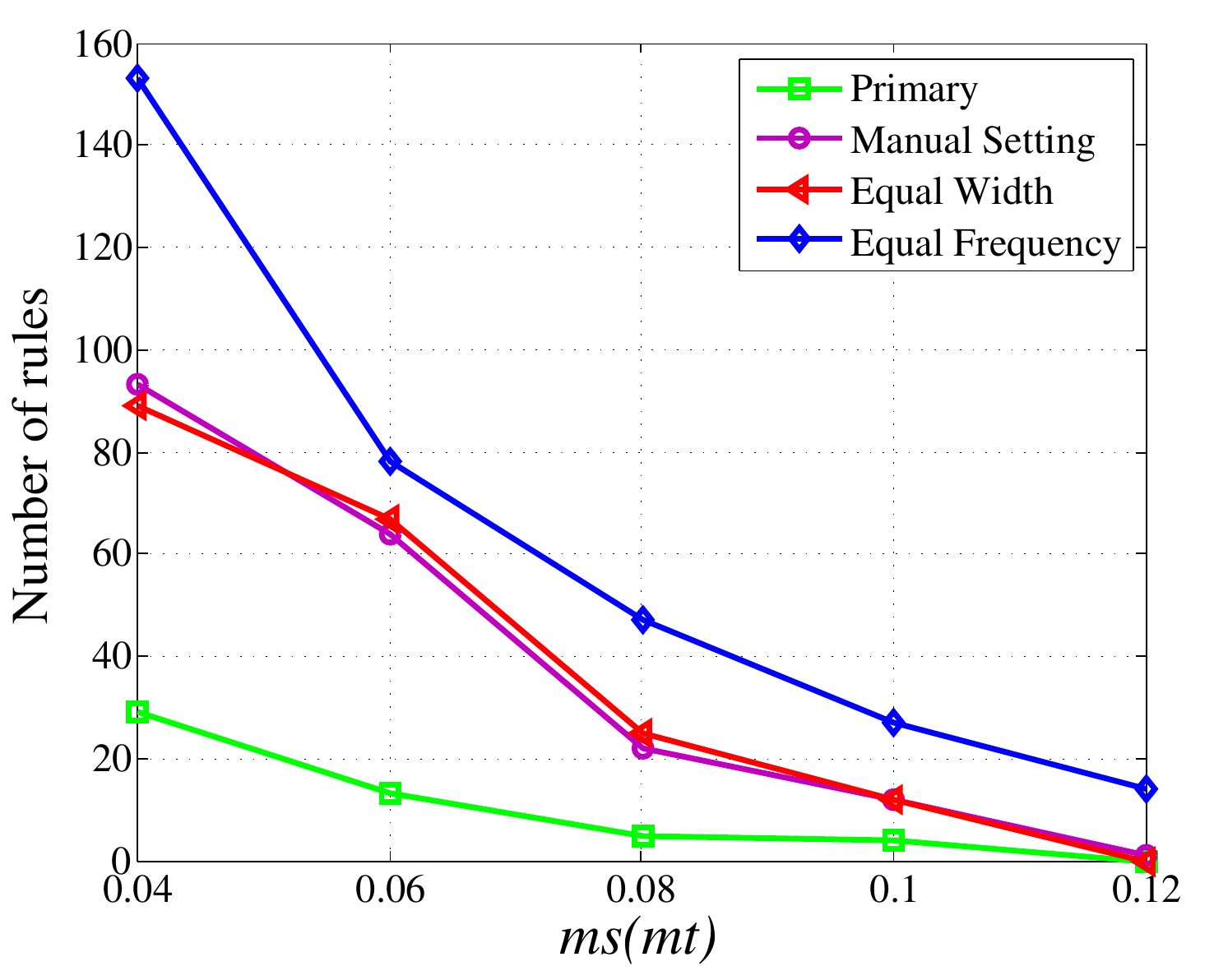}
            \end{minipage}
            }
            \caption{Number of rules: (a) interval number $k = 4$; (b) interval number $k = 8$.}
            \label{figure: rules-comp}
\end{figure}


The evaluation of discretization approaches was performed using the number of generated rules and candidates.
We compare the Equal Width approach, the Equal Frequency approach, the manual discretization setting and primary data which is without discretization.
Let $mc=0.15$, $tc=0.17$ and $ms = mt \in \{$0.04, 0.06, 0.08, 0.10, 0.12$\}$.
Suppose $k$ is the number of intervals.
We set $k = 4$ and $k = 8$ for rule mining, respectively.
We compare the number of candidates and rules, as shown in Figures \ref{figure: candidates-comp}, \ref{figure: rules-comp}.

Figures \ref{figure: candidates-comp}, \ref{figure: rules-comp} show that all discrete approaches can help to mine more candidates and rules from discreted data than not do it from primary data, and the Equal Frequency mine the most.
When $ms = mt = 0.12$, the Equal Frequency can still mine rules, but the others cannot mine any.

We compare the Equal Width approach and the manual discretization setting.
When $k = 4$, the number of candidates and rules of the Equal Width approach and the manual discretization setting have big different, the reason is that a interval may divide into some intervals, which have affects on the number of generated rules. For example, the Equal Width approach obtain a interval [1979, 1998], which includes 1980s and 1990s.
Specifically, when $k = 8$, the number of candidates and rules of them is very similar, the reason is that each interval of them is very similar.

  %
  %
  \subsubsection{The semantically richer}\label{subsubsection: rule-meaning}
We obtain some strong rules using Equal Width and Equal Frequency.
Here we set interval number $k = 4$, $ms = mt = 0.06$, $mc = 0.15$, and $tc = 0.17$.
43 and 68 granular association rules are respectively obtained by Equal Width and Equal Frequency. We respectively list 4 rules of them below.

\noindent The Equal Width approach:

\noindent(Rule 6) $\langle\textrm{age£º [7,24)}\rangle$\\
\indent$\Rightarrow\langle\textrm{genre: action}\rangle$

\noindent(Rule 7) $\langle\textrm{age£º [7,24)}\rangle \wedge \langle\textrm{gender: male} \rangle$\\
\indent$\Rightarrow\langle\textrm{genre: action}\rangle$

\noindent(Rule 8) $\langle\textrm{age£º [7,24)}\rangle \wedge \langle\textrm{gender: male} \rangle$\\
\indent$\Rightarrow \langle\textrm{releaseYear: [1979,1998]}\rangle \wedge \langle\textrm{genre: action}\rangle$

\noindent(Rule 9) $\langle\textrm{age£º [7,24)}\rangle \wedge \langle\textrm{gender: male}\rangle \wedge \langle\textrm{occupation: student}\rangle$\\
\indent$\Rightarrow \langle\textrm{releaseYear: [1979,1998]}\rangle \wedge \langle\textrm{genre: action}\rangle$

\noindent The Equal Frequency approach:

\noindent(Rule 10) $\langle\textrm{age£º [7,25)}\rangle$\\
\indent$\Rightarrow\langle\textrm{genre: action}\rangle$

\noindent(Rule 11) $\langle\textrm{age£º [7,25)}\rangle \wedge \langle\textrm{occupation: student}\rangle$\\
\indent$\Rightarrow \langle\textrm{genre: action}\rangle$

\noindent(Rule 12) $\langle\textrm{age£º [25,31)}\rangle \wedge \langle\textrm{gender: male} \rangle$\\
\indent$\Rightarrow \langle\textrm{releaseYear: [1992,1995]}\rangle \wedge \langle\textrm{genre: comedy}\rangle$

\noindent(Rule 13) $\langle\textrm{age£º [7,25)}\rangle \wedge \langle\textrm{gender: male}\rangle \wedge \langle\textrm{occupation: student}\rangle$\\
\indent$\Rightarrow \langle\textrm{releaseYear: [1992,1995]}\rangle \wedge \langle\textrm{genre: comedy}\rangle$

All rules are quite meaningful from different discrete approaches, and they might be applied to movie recommendation directly.
For Rule 6 indicates that user whose age range from 7 to 24 rate action movies.
We observe that Rule 7 and Rule 8 is finer than Rule 6, which is in turn semantically richer than Rule 6.
Rule 9 obtains the semantically richest rule.
For Rule 11 indicates that user whose age range from 7 to 25 rate action movies.
We observe that Rule 11 is finer than Rule 10, it is similar to the above. Rule 12 mine user age range 25 to 31, but not range 7 to 25, and Rule 13 mine movie genre is comedy but not action, those rules cannot be comparable with Rule 11, but still useful.

  %
  %
  \subsubsection{The setting of interval numbers}\label{subsubsection: intervals-number}
The setting of interval numbers is a key issue in discretization approaches,
so we compare different settings through experiment.
We introduce two parameters $k_1, k_2$, $~k_1$ is number of interval for the numeric data of \textit{User}, $~k_2$ is number of interval for the numeric data of \textit{Movie}.

\begin{figure}[tb]
            \subfigure[]{
            \begin{minipage}[b]{2.7in}
            \centering
             \includegraphics[width=2.65in]{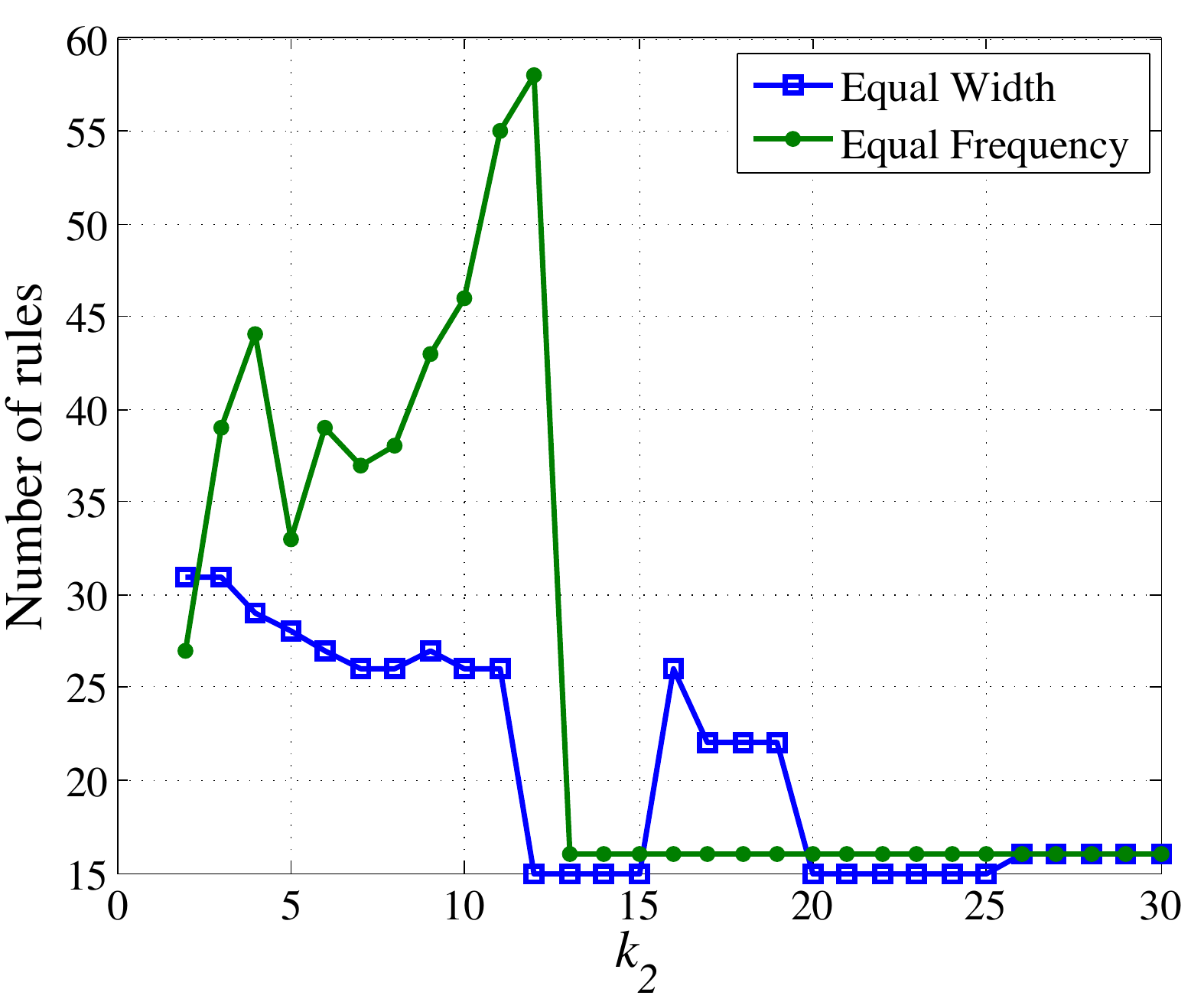}
            \end{minipage}
            }
            \subfigure[]{
            \begin{minipage}[b]{2.7in}
            \centering
             \includegraphics[width=2.65in]{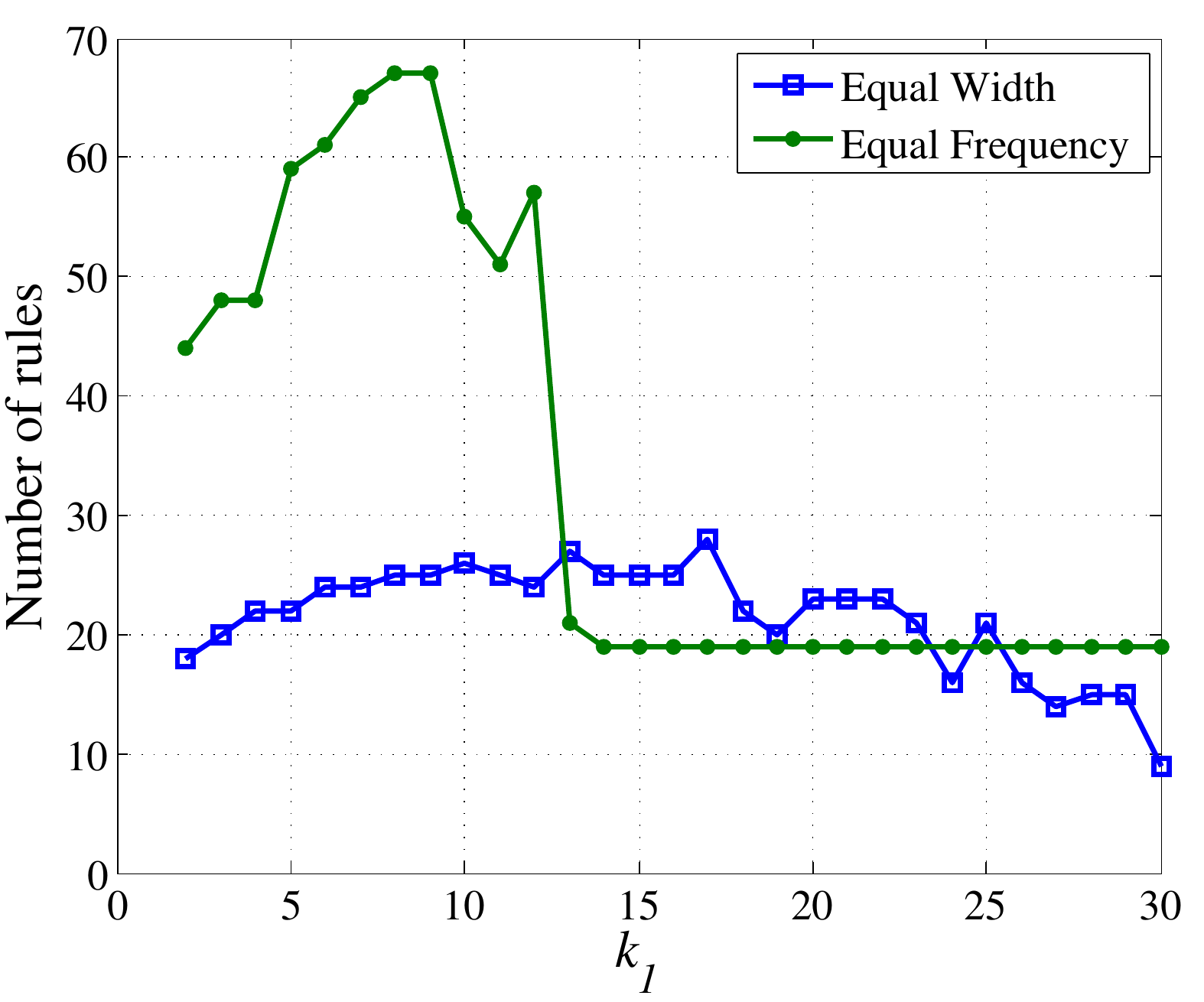}
            \end{minipage}
            }
            \caption{Number of rules: (a) $k_1 = 10$; (b) $k_2 = 11$.}
            \label{figure: check-appropriate-k}
\end{figure}

\begin{figure}[tb]
    \begin{center}
    \includegraphics[width=4in]{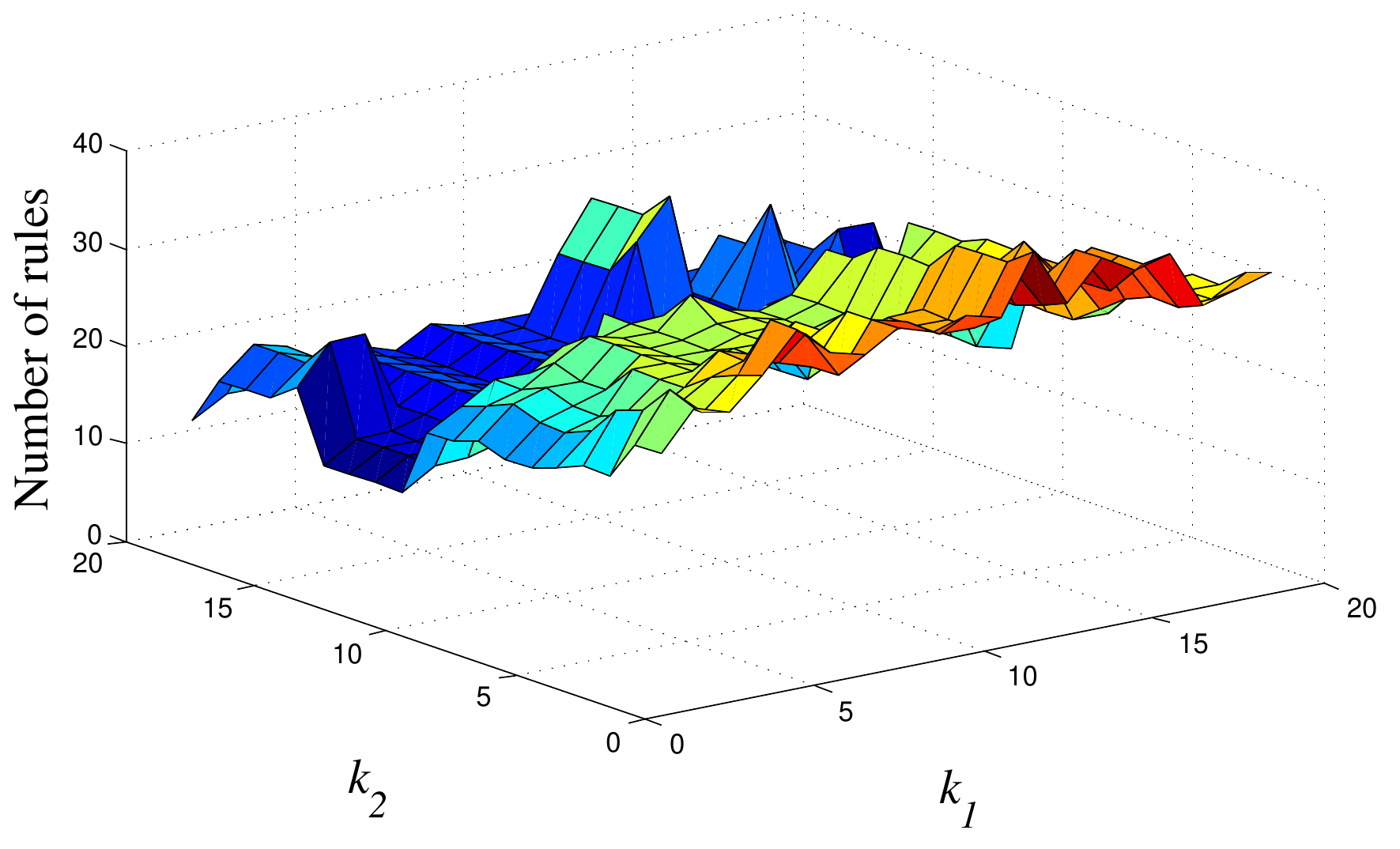}
    \caption{Different settings of interval numbers obtain number of rules through the Equal Width approach}
    \label{figure: Equal-Width-rules}
    \end{center}
\end{figure}

\begin{figure}[tb]
    \begin{center}
    \includegraphics[width=4in]{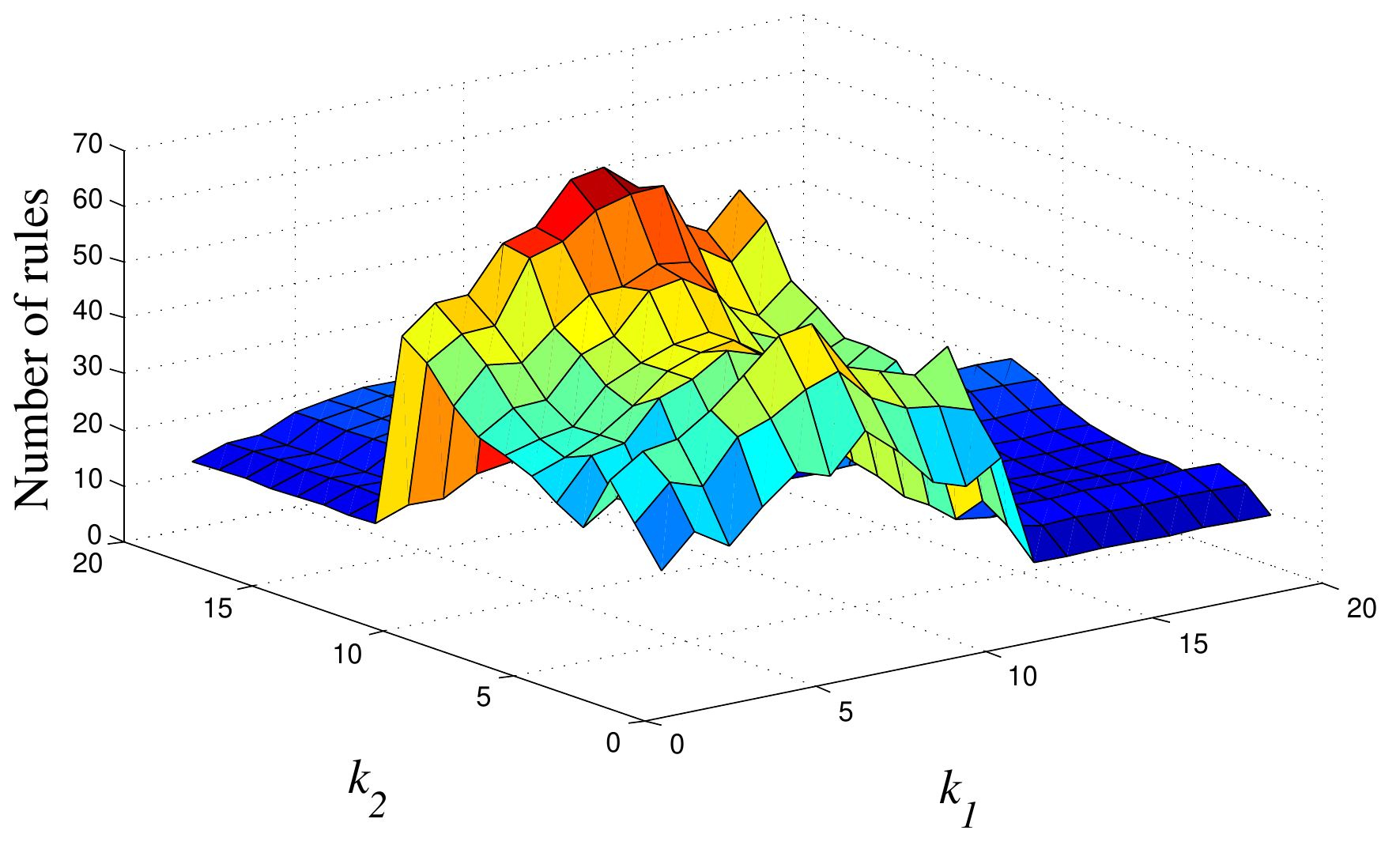}
    \caption{Different settings of interval numbers obtain number of rules through the Equal Frequency approach}
    \label{figure: Equal-Frequency-rules}
    \end{center}
\end{figure}

We set $ms = mt = 0.08$, $mc = 0.15$, and $tc = 0.17$. Firstly we let $k_1 =10$ and let $k_2$ increases from 2 to 30, the number of rules are compared, as depicted in Figure \ref{figure: check-appropriate-k}(a). Secondly we let $k_2 =11$ and let $k_1$ increases from 2 to 30, the number of rules are compared, as drew in Figure \ref{figure: check-appropriate-k}(b). Thirdly we let $k_1$, $k_2$ increase from 2 to 20, respectively, and obtain the corresponding to number of rules, we draw a three-dimensional figure, as shown in Figures \ref{figure: Equal-Width-rules} and  \ref{figure: Equal-Frequency-rules}.

Figure \ref{figure: check-appropriate-k}(a) shows the number of rules decreases as $k_2$ increases, the reason is more interval numbers and lower coverage of intervals, some rules do not satisfy $mt$ that we cannot mine them.
The Equal Frequency approach can mine much more rules than the Equal Width approach at begin.
This is because the number of the users and the movies are well-distributed in the intervals divided by the Equal Frequency approach, more and more intervals can satisfy $mt$ that we can mine much more rules.
When $k_2=12$ of Equal Width and $k_2=13$ of Equal Frequency, the number of rules slumps,
the reason is some rules do not satisfy $mt$. For example, when $k_2=12$, the number of candidates is $18 \times 15$,
while $k_2=13$, the number of candidates is only $18 \times 3$, which is much less.
Finally, the number of rules remains unchanged, because only these rules can be mined before $k_2=30$.

Figure \ref{figure: check-appropriate-k}(b) also shows the number of rules decreases as $k_1$ increases, this is because more interval numbers and lower coverage of intervals, some rules do not satisfy $ms$ that we cannot mine them.
The Equal Frequency approach can mine much more rules when $k_1$ is between 2 and 12.
For the Equal Frequency approach, when $k_1=13$, the number of rules slumps. This is because some rules do not satisfy $ms$.
For instance, when $k_2=12$, the number of candidates is $19 \times 14$,
while $k_2=13$, the number of candidates is only $8 \times 14$, which is much less.
Between $k_1=14$ and $k_1=30$, the number of rules remains unchanged, this is because only these rules can be mined.
For the Equal Width approach, it decreases stable as $k_1$ increases.

Figures \ref{figure: Equal-Width-rules} and  \ref{figure: Equal-Frequency-rules} indicate the number of rules changes with $k_1$ and $k_2$ increase. The Equal Frequency approach can mine more rules than Equal Width. For Figure \ref{figure: Equal-Width-rules}, while $k_1$ range from 10 to 13 and $k_2$ range from 9 to 11, we can obtain more rules. For Figure \ref{figure: Equal-Frequency-rules}, while $k_1$ range from 8 to 10 and $k_2$ range from 10 to 12, we can obtain more rules.
Compare those two Figures, we observe that Figure \ref{figure: Equal-Frequency-rules} is more intuitive than Figure \ref{figure: Equal-Width-rules}.

  %
  %
  \subsection{Discussions}\label{subsection: discussions}
Now we can answer the questions proposed at the beginning of this section.
\begin{enumerate}
\item{
Discretization is an effective preprocessing technique in mining stronger rules, so it outperforms the primary data.
Compared with the manual discretization setting to mine rules and the Equal Width approach, the Equal Frequency approach generates more candidates number and stronger rules.}
\item{
Through discretization, we can obtain much semantically richer rules.}
\item{
   When setting $k_1$ range from 8 to 10 and $k_2$ range from 10 to 12 for Equal Frequency, we obtain certain settings of discrete interval numbers.}
\end{enumerate}

  %
  %
  \section{Conclusions and further works}\label{section: conclusion}
In this paper, we introduced an evaluation and comparison of discretization approaches for granular association rule mining.
With the help of discretization, we mined semantically richer and stronger rules. The Equal Frequency approach
helped generating more rules than the Equal Width approach.
We obtained certain settings of discrete interval to mine much more rules through different approaches.

The following research topics deserve further investigation:
\begin{enumerate}
\item{Preferable discretization approaches. In this work we adopt the Equal Width approach and the Equal Frequency approach. In fact, there are a lot of discretization approaches.
    Many approaches such as rough sets and decision trees would work better on discretized data \cite{Zhu07Generalized,Zhu09RelationshipAmong,Zhu09RelationshipBetween,MinZhu12ACompetition}. We will try to choose some suitable discretization approaches, and design a more appropriate one for granular association rule mining.
}
\item{Intelligent choice. In practice, some data sets contain different numeric data of different attributes, and we use the same discretization approaches to deal with them. However, different algorithms adapt to different data, so that we try to group different algorithms to realize intelligent choice for discretization of the same data set. The improved scheme is more valuable in practical application.
}

\end{enumerate}

  %
  %
  \section*{Acknowledgements}\label{section: acknowledgements}
This work is in part supported by National Science Foundation of China under Grant No. 61170128, the Natural Science Foundation of Fujian Province, China under Grant Nos. 2011J01374, 2012J01294, and Fujian Province Foundation of Higher Education under Grant No. JK2012028.

  %
  %
\bibliographystyle{elsart-num-sort}

\end{document}